\documentclass[aps, superscriptaddress, showpacs, floatfix, twocolumn]{revtex4-1}
\usepackage{amsmath}
\usepackage{amssymb}
\usepackage{hyperref}
\usepackage{graphicx}
\begin{document}
\title{Entanglement entropy and computational complexity of the Anderson impurity\\ model out of equilibrium I: quench dynamics}
\author{Zhuoran He}
\affiliation{Department of Physics, Columbia University, New York, New York 10027, USA}
\author{Andrew J. Millis}
\affiliation{Department of Physics, Columbia University, New York, New York 10027, USA}

\begin{abstract}
\vspace*{0\baselineskip}
We study the growth of entanglement entropy in density matrix renormalization group calculations of the real-time quench dynamics of the Anderson impurity model. We find that with appropriate choice of basis, the entropy growth is logarithmic in both the interacting and noninteracting single-impurity models. The logarithmic entropy growth is understood from a noninteracting chain model as a critical behavior separating regimes of linear growth and saturation of entropy, corresponding respectively to an overlapping and gapped energy spectra of the set of bath states. We find that with an appropriate choices of basis (energy-ordered bath orbitals), logarithmic entropy growth is the generic behavior of quenched impurity models. A noninteracting calculation of a double-impurity Anderson model supports the conclusion in the multi-impurity case. The logarithmic growth of entanglement entropy enables studies of quench dynamics to very long times.
\end{abstract}

\maketitle

\section{Introduction}
It has been a long-standing challenge to develop efficient real-time impurity solvers to study the dynamics of a system out of equilibrium. The Anderson impurity model \cite{Anderson61} (AIM), a single spin-degenerate orbital with an intra-orbital Hubbard interaction $U$ coupled to a bath of noninteracting orbitals, is of fundamental importance in its own right as a nontrivial but solvable \cite{Wiegmann83,Tsvelick83} interacting electron model and as an auxiliary problem for dynamical mean field theory \cite{Georges96,Kotliar06}. The nonequilibrium properties of this model \cite{Aoki14,Gramsch13} provide an important laboratory for the development of real-time methods \cite{Hettler98,Cohen15,Balzer15,Kreula16}.

The density matrix renormalization group (DMRG) \cite{Schollwock05} is a powerful numerical method for solving low-dimensional electron problems including the Anderson impurity model. In DMRG, the wave function of the system is represented by a matrix product state (MPS). Every matrix in the MPS corresponds to a local degree of freedom. If these degrees of freedom are labeled by an index $n$, then the logarithm of the bond dimension $D_n$ between the $n$-th and $(n+1)$-th matrices of the MPS is greater than or of the order of the entanglement entropy $S_n$ between orbitals $1,2\ldots n$ and $n+1\ldots N$. The rate of growth with time of the entanglement entropy across the maximum entropy cut controls the maximum time achievable in a real-time DMRG calculation.

In this paper we study the growth of entanglement entropy in DMRG calculations, focusing primarily on the single-impurity Anderson model. Our work is motivated by the recent results of Wolf \textit{et~al.}~\cite{Wolf14}, who found that a particular ``star geometry'' arrangement of bath orbitals provided a favorable growth of entanglement entropy. We consider a variety of bath Hamiltonians and find that the growth of entanglement entropy $S_n$ across the cut at $n$ is controlled by whether the spectra of MPS orbitals $1,2\ldots n$ overlaps that of the $n+1\ldots N$ orbitals. The star geometry with energy-ordered bath orbital arrangement in the MPS ensures no energy overlap at every bond, leading to a logarithmic growth of maximum entanglement entropy with time. This method of controlling entropy growth using energy separation is hypothesized to work for multi-impurity models as well; the hypothesis is supported by a calculation of a non-interacting double-impurity Anderson model.

The rest of the paper is organized as follows. Section \ref{sec:theory} describes the single-impurity Anderson model (SIAM) and Sec.~\ref{sec:method} describes our implementation of DMRG, which differs slightly from the standard implementation. In Secs.~\ref{sec:results}--\ref{sec:double-impurity}, we show results obtained using our method for the SIAM, discuss the logarithmic growth of entropy and also present a few results for the quenched double-impurity Anderson model. Section \ref{sec:conclusion} is a conclusion and summary.

\section{Theory \label{sec:theory}}
\begin{figure}
\includegraphics[width=0.8\columnwidth]{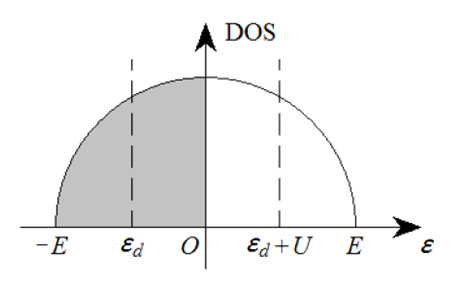}
\caption{The density of states of the bath orbitals. We consider a semicircle DOS with a half band width $E$. The bath is initially half-filled, and the $d$-orbital energy $\epsilon_d$ and $\epsilon_d+U$ are symmetric about the Fermi level at $0$.}
\label{fig:DOS}
\end{figure}

We focus on the single-impurity Anderson model (SIAM) defined by the Hamiltonian
\vspace{-1.5ex}
\begin{align}
&H=H_d+H_\mathrm{bath}+H_\mathrm{mix},\phantom{\frac{1}{2}}
\label{eq:H-1}\\
&H_d=\sum_\sigma\epsilon_d\,d_\sigma^\dagger d_\sigma+U\,d_\uparrow^\dagger d_\uparrow d_\downarrow^\dagger d_\downarrow,
\label{eq:H-2}\\
&H_\mathrm{bath}=\sum_{k\sigma}\epsilon_k\,c_{k\sigma}^\dagger c_{k\sigma},
\label{eq:H-3}\\
&H_\mathrm{mix}=\sum_{k\sigma}V_k\,d_\sigma^\dagger c_{k\sigma}+\mathrm{h.c.}.
\label{eq:H-4}
\end{align}
Here $k$ labels the $\mathcal{N}\rightarrow\infty$ bath orbitals and $\sigma$ labels the spin. For simplicity we take the impurity-bath coupling amplitudes $V_k=V/\sqrt{\mathcal{N}}$ to be $k$-independent. We define the bath density of states $\mathrm{DOS}(\epsilon)=\frac{1}{\mathcal{N}}\sum_k\delta(\epsilon-\epsilon_k)$ and consider a semicircle with a half band width $E$. The initial state that we consider is a product state
\begin{align}
|\Psi_{t=0}\rangle=|\Psi_0\rangle_d\otimes|\mathrm{FS}\rangle_\mathrm{bath},
\label{eq:pure-initial-state}
\end{align}
where the Fermi-sea state $|\mathrm{FS}\rangle_\mathrm{bath}$ of the bath is initially half-filled. In the studies we present here, the $d$-orbital energies $\epsilon_d$ and $\epsilon_d+U$ are chosen to be symmetric about the Fermi level at $0$. This is not an essential requirement of our theory. The situation is depicted in Fig.~\ref{fig:DOS}. The formalism is easily generalized to a mixed state with an initial density matrix $\rho_{t=0}=(\rho_0)_d\otimes(\rho_0)_\mathrm{bath}$, but here we will focus on a pure initial state.

The numerical methods we use require a truncation of the bath to a finite number $N$ of bath orbitals. To choose the best truncation, we first calculate the hybridization function $\Delta_\sigma(t,t')=-i\sum_k|V_k|^2\langle\mathcal{T_C}\,c_{k\sigma}(t)c_{k\sigma}^\dagger (t')\rangle_\mathrm{bath}$. Here we work with the Keldysh-contour Green's function $-i\langle\mathcal{T_C}\ldots\rangle_\mathrm{bath}$, where operators $c_{k\sigma}(t)$ and $c_{k\sigma}^\dagger (t')$ are in the Heisenberg picture evolving via $H_\mathrm{bath}$ and the mean value $\langle\ldots\rangle_\mathrm{bath}$ is taken with respect to $|\mathrm{FS}\rangle_\mathrm{bath}$. With our choice of the semicircular DOS and constant $V_k$, the hybridization function can be evaluated at $\mathcal{N}\rightarrow\infty$ as
\begin{align}
\Delta_\sigma(t,t')=\left\{
\begin{array}{lr}
\displaystyle
-\frac{V^2}{E\tau}[H_1(E\tau)+iJ_1(E\tau)], & t\succ_\mathcal{C}t',\\
\vspace{-1ex} & \ \\
\displaystyle
-\frac{V^2}{E\tau}[H_1(E\tau)-iJ_1(E\tau)], & t\prec_\mathcal{C}t',
\end{array}
\right.
\label{eq:hyb-func-analytic}
\end{align}
where $\tau=t-t'$, $H_1$ is the 1st-order Struve function and $J_1$ is the 1st-order Bessel function. The symbols $\succ_\mathcal{C}$ and $\prec_\mathcal{C}$ refer to Keldysh-contour ordering. Then we fit the hybridization function to that of a finite bath with only $N$ orbitals, i.e.,
\begin{align}
\Delta_\sigma(t,t')&\approx-i\sum_{j=1}^N V_j^2\langle\mathcal{T_C}\,c_{j\sigma}(t)c_{j\sigma}^\dagger(t')\rangle_\mathrm{bath}\nonumber\\
&=-i\sum_{j=1}^N V_j^2\left[\Theta_\mathcal{C}(t,t')-n_{j\sigma}^0\right]\!e^{-i\epsilon_j(t-t')}.
\label{eq:hyb-func-fit}
\end{align}
In fitting Eq.~\eqref{eq:hyb-func-analytic} with Eq.~\eqref{eq:hyb-func-fit}, all $2N$ real parameters $\epsilon_j$ and $V_j$ are varied to minimize the least-square error. The occupancies $n_{j\sigma}^0$ are chosen to be either $0$ or $1$ to fit the $t\succ_\mathcal{C}t'$ and $t\prec_\mathrm{C}t'$ parts independently and to make the initial state of the finite bath a Slater determinant. This is possible even if the original bath was at nonzero temperature. Since our bath is particle-hole symmetric, we choose $N$ to be even to preserve this symmetry.

The number $N$ of bath orbitals controls the maximum time $t_N\lesssim 2\pi N/|E_\mathrm{max}-E_\mathrm{min}|=\pi N/E$ up to which the exact hybridization function is reproduced with good accuracy. For example, $N=40$ bath orbitals are enough to reach $Et\lesssim 100$ and $N=170$ orbitals can reach $Et\lesssim 500$. Adding more orbitals increases the maximal time that can be reached, but does not significantly improve the accuracy of the fit at shorter times.

\section{Method \label{sec:method}}
We use DMRG/MPS methods to carry out the time evolution. We represent the SIAM wave function $|\Psi(t)\rangle$ as an entangled state between the impurity $d$ orbital and the bath, i.e.,
\begin{align}
|\Psi(t)\rangle=\sum_i\,&c_i(t)|i\rangle_d\otimes|\Psi_i(t)\rangle_\mathrm{bath},
\label{eq:4-MPS}
\end{align}
where $i$ sums over the 4 impurity states $|0\rangle$, $\left|\uparrow\right>$, $\left|\downarrow\right>$ and $\left|\uparrow\downarrow\right>$. Every bath state $|\Psi_i(t)\rangle_\mathrm{bath}$ is a normalized matrix product state (MPS), with the $c_i(t)$'s being the normalizing coefficients. Eq.~\eqref{eq:4-MPS} is a Schmidt decomposition of $|\Psi(t)\rangle$ between the $d$ orbital and the bath if $|\Psi(t)\rangle$ is a simultaneous eigenstate of $N_\uparrow$ and $N_\downarrow$, the total numbers of spin-up and spin-down electrons. This representation differs from the conventional DMRG in that it removes the $d$ orbital from the MPS, enabling analysis of the entanglement among the bath orbitals.

We evolve the wave function $|\Psi(t)\rangle$ using the interaction picture of $H_0=H_d+H_\mathrm{bath}$. The wave function evolves according to
\begin{align}
|\Psi(t)\rangle=\mathcal{T}e^{-i\int_0^t dt' \hat{H}_\mathrm{mix}(t')}|\Psi_{t=0}\rangle,
\label{eq:psi-t}
\end{align}
where $\mathcal{T}$ is the time-ordering symbol and
\vspace{-0.5ex}
\begin{align}
&\quad\;\hat{H}_\mathrm{mix}(t)=e^{iH_0t}H_\mathrm{mix}\,e^{-iH_0t}\phantom{\frac{1}{2}}\nonumber\\
&=\sum_{j\sigma}V_j\,e^{i(Un_{d\bar{\sigma}}+\epsilon_d-\epsilon_j)t}d_\sigma^\dagger c_{j\sigma}+\mathrm{h.c.},
\end{align}
where $\bar{\sigma}$ is the opposite spin of $\sigma$. The main advantage of the interaction picture is that $\hat{H}_\mathrm{mix}(t)$ typically has a narrower spectral radius than $H_0$ (bath bandwidth $\sim E$ large compared with impurity level width $\sim V^2/E$).

We evaluate Eq.~\eqref{eq:psi-t} by discretizing the time interval into time steps $\Delta t$. The operator at the time step centered on time $t$ is
\begin{align}
\tilde{H}_\mathrm{mix}(t)&=\frac{1}{\Delta t}\int_{t-\Delta t/2}^{t+\Delta t/2}\hat{H}_\mathrm{mix}(t')\,dt'
\phantom{\sum\sum}
\nonumber\\
&=\sum_{j\sigma}\tilde{V}_{j\sigma}(t)d_\sigma^\dagger c_{j\sigma}+\mathrm{h.c.},
\phantom{\int}
\label{eq:H-eff}
\end{align}
with the coupling amplitudes
\begin{align}
\tilde{V}_{j\sigma}(t)=V_j\,e^{i(Un_{d\bar{\sigma}}+\epsilon_d-\epsilon_j)t}\,\mathrm{sinc}(\textstyle\frac{Un_{d\bar{\sigma}}+\epsilon_d-\epsilon_j}{2}\Delta t).
\end{align}
The errors of both the mid-point Hamiltonian $\hat{H}_\mathrm{mix}(t)$ and the time-averaged Hamiltonian $\tilde{H}_\mathrm{mix}(t)$ are $\mathcal{O}(\Delta t^2)$. The latter choice is preferred if the bath bandwidth is large compared with the level width, because the very high and very low-energy bath orbitals are suppressed by the $\mathrm{sinc}$ function.

To apply the Hamiltonian $\tilde{H}_\mathrm{mix}(t)$ to the wave function $|\Psi(t)\rangle$ in Eq.~\eqref{eq:4-MPS}, we work in the Jordan-Wigner transformed representation with the $d$ orbital being the first orbital ($d$ and $d^\dagger$ having no Jordan-Wigner signs). The Hamiltonian in Eq.~\eqref{eq:H-eff} is rewritten as
\begin{align}
&\tilde{H}_\mathrm{mix}(t)=\sum_\sigma(-1)^{n_{d\bar{\sigma}}}d_\sigma^\dagger\tilde{c}_\sigma(t)+\mathrm{h.c.},\\
&\tilde{c}_\sigma(t)=\sum_j\tilde{V}_{j\sigma}(t)\,(-1)^{n_1+\cdots+n_{j\!-\!1}}\tilde{c}_{j\sigma},
\label{eq:bath-ops}
\end{align}
where the $\tilde{c}_{j\sigma}$ is the Jordan-Wigner transformed $c_{j\sigma}$. The two operators are related by
\begin{align}
c_{j\sigma}=(-1)^{n_d+n_1+\cdots+n_{j-1}}\tilde{c}_{j\sigma},
\end{align}
so that the operators $\tilde{c}_{j\sigma}$ and $\tilde{c}_{j'\sigma'}$ for $j\neq j'$ commute. We do the same Jordan-Wigner transform the two spins of the same orbital, so that $\tilde{c}_{j\uparrow}$ and $\tilde{c}_{j\downarrow}$ still anticommute. But this is easy to handle with a local $4\times 4$ matrix. The bath operator $\tilde{c}_\sigma(t)$ in Eq.~\eqref{eq:bath-ops} is then represented by a matrix-product operator (MPO)
\begin{align}
\tilde{c}_\sigma(t)=\left[0,1\right]
\prod_{j=1}^{N}\begin{bmatrix}
I & 0\\
\tilde{V}_{j\sigma}(t)\tilde{c}_{j\sigma} & (-1)^{n_j}
\end{bmatrix}\begin{bmatrix}
1\\0
\end{bmatrix},
\end{align}
where the $j=1$ matrix is left-multiplied by $[0,1]$ to pick the second row, and the $j=N$ matrix is right-multiplied by $[1,0]^T$ to pick the first column. The MPO has a bond dimension of 2. We can similarly express $\tilde{c}_\sigma^\dagger(t)$ in terms of $\tilde{c}_{j\sigma}^\dagger$. The Hamiltonian $\tilde{H}_\mathrm{mix}(t)$ can then act on $|\Psi(t)\rangle$ following DMRG routines \cite{Schollwock05}.

The final evolution scheme is given by
\begin{align}
|\Psi(t+\Delta t)\rangle\approx e^{-i\tilde{H}_\mathrm{mix}\left(t+\frac{\Delta t}{2}\right)\Delta t\,}|\Psi(t)\rangle
\end{align}
with the exponential factor Taylor expanded into a 4th-order polynomial of $\tilde{H}_\mathrm{mix}(t\!+\!\Delta t/2)$. The narrow spectral radius of $\tilde{H}_\mathrm{mix}\Delta t$ ensures good unitarity of the 4th-order truncation. Since the bath operators $\tilde{c}_\sigma$ and $\tilde{c}_\sigma^\dagger$ are long-range, we cannot locally exponentiate the Hamiltonian following the time-evolving block decimation (TEBD) \cite{Vidal03} method. We adjust the truncation error tolerance of the singular value decomposition (SVD) in DMRG routines according to the MPS norm so that higher-order terms do not take much time to calculate. We also parallelize the calculations of the 4 MPSs in Eq.~\eqref{eq:4-MPS} on 4 cores and use total numbers of spin-up and spin-down electrons as symmetries to speed up the calculation.

\section{Results \label{sec:results}\label{sec:results-physics}}

\begin{figure}[b]
\includegraphics[width=0.9\columnwidth]{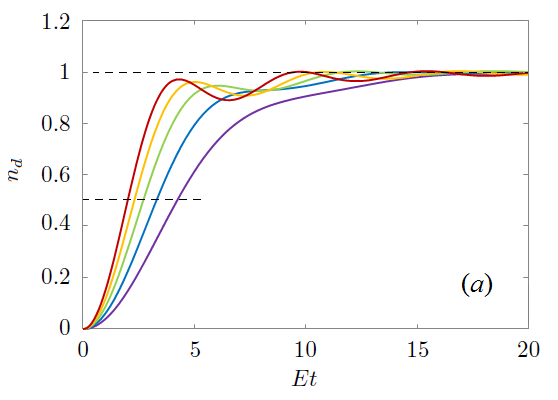}
\includegraphics[width=0.9\columnwidth]{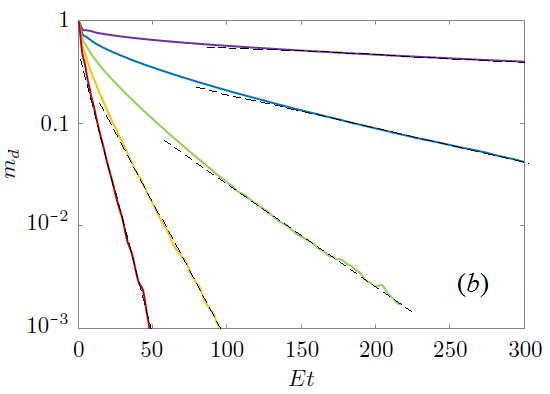}
\caption{(Color online) The charge and spin dynamics of the SIAM. (a) The occupancy $n_d$ v.s.~$t$ starting from $|\Psi_0\rangle_d=|0\rangle_d$ with impurity-bath coupling $V/E=0.2$, $0.25$, $0.3$, $0.35$, $0.4$ from bottom to top; (b) The magnetic moment $m_d$ v.s.~$t$ starting from $|\Psi_0\rangle_d=\left|\uparrow\right>_d$ with the same values of $V/E$ from top to bottom. Dashed lines show the linear fits used to obtain the long-time relaxation rates in Fig.~\ref{fig:spin-relax}b. Hubbard $U/E=1$. The number of bath orbitals we used was $N=20$ in (a) and $N=130$ in (b).\label{fig:nd}}
\end{figure}

In this section we show some results obtained for the interacting SIAM with $U/E=1$ using the method and other model parameters described in Secs.~\ref{sec:theory} and \ref{sec:method}. The impurity-bath coupling $V/E=0.1\sim 0.5$. This is the parameter range of interest. The impurity level width $\sim V^2/E$ remains smaller than the band width $\sim E$ while the Kondo temperature $T_K\approx 0.4Ve^{-\pi E^2/16V^2}$ (from \cite{Wang08}) can change by orders of magnitudes.

\begin{figure}[b]
\includegraphics[width=0.9\columnwidth]{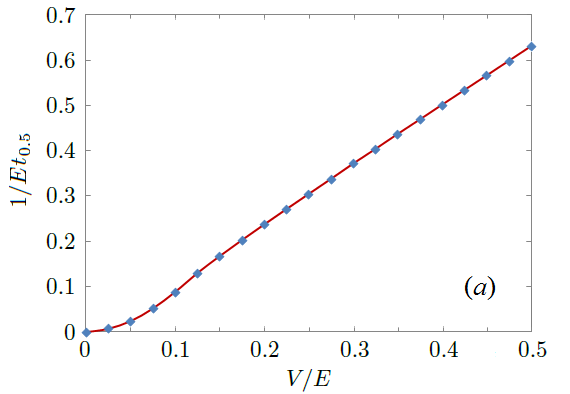}
\includegraphics[width=0.9\columnwidth]{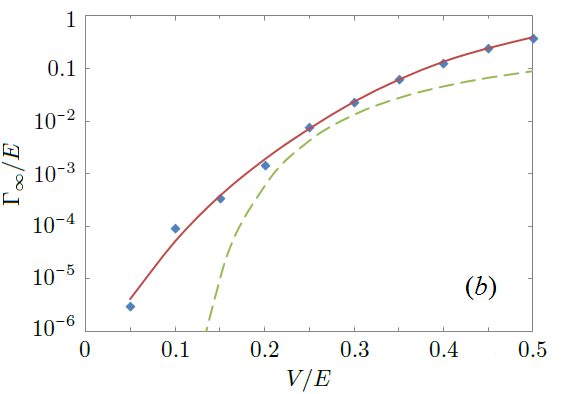}
\caption{(Color online) The charge equilibration rate $1/t_{0.5}$ in (a) and the spin relxation rate $\Gamma_\infty\equiv d\ln m_d/dt|_{t\rightarrow\infty}$ in (b) obtained from $n_d(t)$ and $m_d(t)$ (partly shown in Figs.~\ref{fig:nd}a and \ref{fig:nd}b). $\Gamma_\infty$ is estimated using $m_d(t)$ up to $Et\leq 600$. Hubbard interaction $U/E=1$.
\label{fig:spin-relax}}
\end{figure}

Fig.~\ref{fig:nd}a shows the charge relaxation dynamics, obtained by starting from an initially empty $d$ orbital $|\Psi_0\rangle_d=|0\rangle_d$ and a half-filled Fermi-sea state $|\mathrm{FS}\rangle_\mathrm{bath}$ for the bath. Our choice of particle-hole symmetric parameters ensures that $n_d=\langle n_{d\uparrow}\rangle+\langle n_{d\downarrow}\rangle$ always equilibrates to $1$ so long as the impurity-bath coupling $V$ is not big enough to form a bound state on the impurity. We see in agreement with previous work \cite{Eckstein09,Wolf14} that the charge equilibration proceeds relatively rapidly. The reciprocal of the time $t_{0.5}$ it takes to reach $n_d=0.5$ is plotted in Fig.~\ref{fig:spin-relax}a. At small $V/E\lesssim 0.1$, $t_{0.5}\sim V^{-2}$ is inversely proportional to the $d$-level width $\sim V^2/E$. For $V/E\gtrsim 0.15$, the rate $1/t_{0.5}$ crosses over to approximately linear in $V$ and the equilibration process in Fig.~\ref{fig:nd}a becomes more oscillatory as we are approaching the formation of a bound state on the impurity. The variation of charge equilibration rates with $V$ can be seen in calculations performed for a noninteracting SIAM, suggesting that the charge relaxation physics is essentially due to hybridization. The Hubbard $U$ does not qualitatively change the behavior of the model.

Fig.~\ref{fig:nd}b shows the spin relaxation dynamics obtained by starting from $\left|\uparrow\right>_d\otimes|\mathrm{FS}\rangle_\mathrm{bath}$, a fully spin-polarized $d$ orbital and the same half-filled Fermi-sea state $|\mathrm{FS}\rangle_\mathrm{bath}$ of the bath. The magnetization $m_d=\langle n_{d\uparrow}\rangle-\langle n_{d\downarrow}\rangle$ relaxes much more slowly than the charge, again in agreement with previous results \cite{Cohen13,Rostami16}. The asymptotic behavior of $m_d$ v.s.~$t$ shows approximately an exponential tail, with the relaxation rate $\Gamma_\infty\equiv d\ln m_d/dt|_{t\rightarrow\infty}$ plotted in Fig.~\ref{fig:spin-relax}b. $\Gamma_\infty$ is estimated by fitting $\ln m_d(t)$ v.s.~$t$ to a straight line for $t_\mathrm{max}/2<t<t_\mathrm{max}$, where $t_\mathrm{max}$ is the maximum time reached in the simulation for the slope. The solid red line is a trend line. We also show as the dashed green line the analytical result --- the Kondo temperature $T_K$ calculated using the formula in \cite{Wang08} and interpreted as a relaxation rate. 

The Kondo result has a similar magnitude and $V$ dependence to the calculated results. The numerical differences at large $V$ arise from relaxation processes associated with valence fluctuations not included in the Kondo limit, while the more pronounced differences at small $V$ are an intermediate asymptotics effect. For small $V$, even at the very long times ($Et\leq 600$) accessible to our method, the magnetization $m_d$ is still substantial, so the Kondo-limit expression, which gives the linear response relaxation for small magnetization ($m_d\rightarrow 0$), is not applicable. Evidently, the nonlinear response (relaxation of a finite $m_d$) is stronger than the linear response. Developing a theory of the relaxation in the small $V$ and intermediate $m_d$ regime is an interesting open question. For intermediate $V/E\simeq 0.25$, the theoretical result is within a factor of $2$ of the numerical one with the differences likely arising from the convention used for the Kondo temperature $T_K$.

\section{Logarithmic growth of entropy \label{sec:results-entropy}}

A remarkable feature of the simulations reported here is the long time scales that can be reached; these time scales are necessary to reveal, for example, the magnetization decay. As we show in this section, this is possible because the maximum entanglement entropy of the 4 bath MPSs in Eq.~\eqref{eq:4-MPS} grows only logarithmically during the simulation, which means the long times are not exponentially hard to reach, but are of only polynomial time complexity.

\subsection{Entanglement entropy growth in SIAM}

\begin{figure}
\includegraphics[width=0.9\columnwidth]{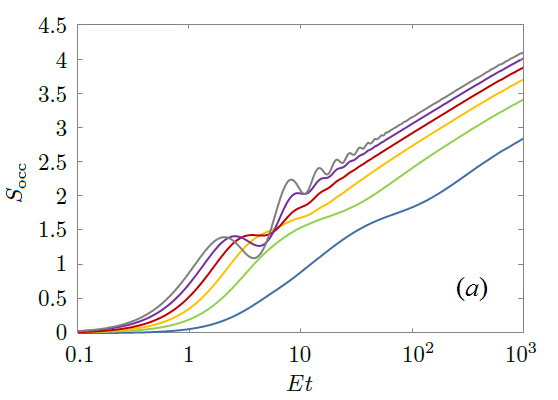}
\includegraphics[width=0.9\columnwidth]{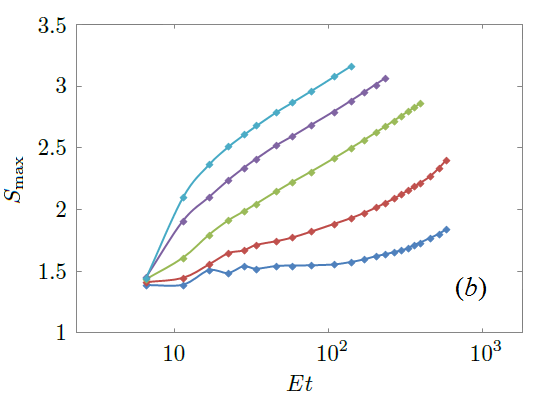}
\caption{(Color online) The logarithmic growth of entanglement entropy. (a) The entanglement entropy of the initially occupied part of the bath with the rest of the system at $U=0$ and $V/E=0.1$, $0.2,\ldots,0.6$ from bottom to top. (b) The maximum entanglement entropy encountered in the interacting SIAM simulation v.s.~time $t$ at $U/E=1$ and $V/E=0.15$, $0.2$, $0.25$, $0.3$, $0.35$ from bottom to top.
\label{fig:entropy}}
\end{figure}

In this section, we compare the maximum entanglement entropy of the interacting SIAM ($U/E=1$) with a noninteracting SIAM ($U=0$) with $\epsilon_d=0$ at the Fermi level. Both models start from the same initial condition $|0\rangle_d\otimes|\mathrm{FS}\rangle_\mathrm{bath}$ with an empty $d$ orbital and a half-filled bath in Fig.~\ref{fig:DOS}. Results of the entanglement entropy are shown in Fig.~\ref{fig:entropy}. The entropy growth starting from a spin-polarized impurity $\left|\uparrow\right>_d\otimes|\mathrm{FS}\rangle_\mathrm{bath}$ is also logarithmic but takes smaller values.

The curves in Figs.~\ref{fig:entropy}a and \ref{fig:entropy}b are obtained in slightly different ways. Fig.~\ref{fig:entropy}a shows results obtained for a noninteracting simulation of $N=1000$ bath orbitals all coupled to one empty $d$ orbital at the Fermi level. We then plot the entanglement entropy between the 500 bath orbitals below the Fermi level with the other 501 orbitals up to $Et=1000$ and the data shows a logarithmic growth of entanglement entropy at all values of the impurity-bath coupling $V$. At long times, the slopes of the curves are the same; only the offset and the transient entropy growth depend on $V$.

Fig.~\ref{fig:entropy}b shows the interacting model ($U/E=1$, $\epsilon_d=-U/2$). At each time, the hybridization function was fitted with the minimal number of bath orbitals needed to obtain a fit with a root-mean-square error (RMSE) of $3\times 10^{-4}$. Notice that the hybridization fit is independent of $U$ and can be done before simulation. We then plot the maximum entanglement entropy $S_\mathrm{max}$ seen on all bonds of the 4 bath MPSs encountered during the simulation from $t=0$ to the maximum simulation time $t_\mathrm{max}$ allowed by the hybridization fit v.s.~$t_\mathrm{max}$. $S_\mathrm{max}$ may be encountered before $t_\mathrm{max}$ due to the finite bath effect. So Fig.~\ref{fig:entropy}b takes into account the possibility of using the finite bath effect to limit entropy growth. But still the logarithmic growth of entropy and the independence of the steady-state slope of $S-\log t$ on the impurity-bath coupling $V$ is the same as the noninteracting SIAM simulation in Fig.~\ref{fig:entropy}a. These two properties mean $S\leq c\ln t$, and therefore the bond dimension $D\sim e^S\leq t^c$, which means the interacting SIAM can be simulated in polynomial time $\mathcal{O}(D^3)=\mathcal{O}(t^{3c})$ of $t$.

\subsection{Analysis of entropy growth \label{sec:results-entropy-analysis}}

\begin{figure}
\includegraphics[width=0.9\columnwidth]{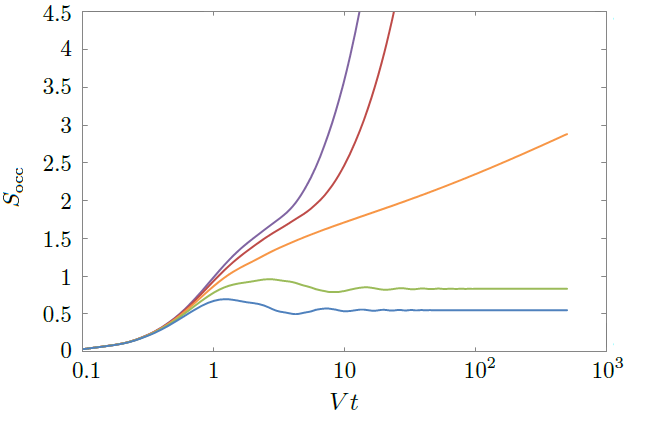}
\includegraphics[width=0.85\columnwidth]{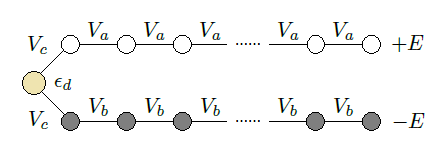}
\includegraphics[width=0.95\columnwidth]{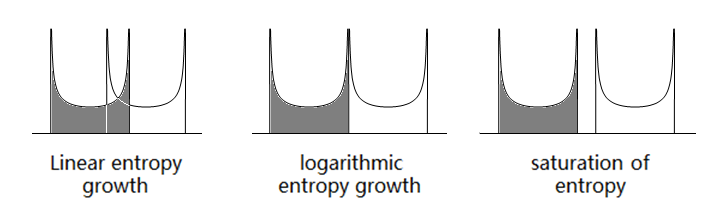}
\caption{(Color online) The two semi-infinite chain model (lower panel) and its critical behavior at $E=V_a+V_b$ (upper panel) with $\epsilon_d=0$, $V_a=V_b\equiv V$, $V_c=0.5V$ and $E/V=1$, $1.5$, $2$, $2.5$, $3$ from top to bottom. $S_\mathrm{occ}$ is the entanglement entropy between chain $b$ (initially occupied) with the rest of the system (initially empty impurity and chain $a$).
\label{fig:chain}}
\end{figure}

To understand the logarithmic growth of entropy, we consider a noninteracting chain model, as is shown in Fig.~\ref{fig:chain}. In this model, the impurity is coupled to two semi-infinite chains. We choose a constant hopping amplitude between the bath sites in each chain. By adjusting the on-site energy difference of the two chains, we can vary the densities of states as shown in the lower panel of Fig.~\ref{fig:chain}, obtaining either overlapping, gapped or just touching spectra. Our computation of the entanglement entropy $S_\mathrm{occ}$ across the impurity site shows that we have linear growth, logarithmic growth, and saturation, respectively. In the numerical test we did, chain $a$ was initially empty and chain $b$ was initially full. But the conclusion is found to hold for randomized initial occupancies, too.

The logarithmic growth of $S_\mathrm{occ}$ is seen at a critical $E=V_a+V_b$, at which the density of states (DOS) of the two semi-infinite chains touch at only one energy point. When $E>V_a+V_b$, the system is gapped and entropy growth saturates. This can be explained by the lack of energy eigenstates that are extended in both regions $a$ and $b$, which then means that particles (or holes) that are originally in $a$ cannot go into $b$ and vice versa beyond a penetration depth determined by the gap, which then puts an upper bound on the entanglement entropy between $a$ and $b$. This energy barrier works for a general initial occupancy. Starting from any product state, so long as the semi-chains $a$ and $b$ are gapped, the entropy must saturate.

When $E<V_a+V_b$, there is a finite overlap of the DOS of the two semi-infinite chains and we see a linear growth of entanglement entropy in Fig.~\ref{fig:chain}. In rare cases this does not happen. For example, for a uniform chain $V_a=V_b=V_c$ and $E=\epsilon_d=0$, the entropy growth is logarithmic rather than linear. But this behavior depends on the initial occupancy. If the occupied sites are randomized, or if the model parameters are slightly modified to deviate from a uniform chain, the expected behavior of a linear growth of the entanglement entropy is seen between $a$ and $b$. The energy criterion guarantees that particles do not enter the forbidden regions of a noninteracting bath. But once the energy barrier is not at work, it is difficult in general, though not impossible, to organize the migrated particles into a low entanglement entropy state to make the MPS matrices small.

The logarithmic growth of entropy in Fig.~\ref{fig:entropy}b can be understood as the result of arranging the bath orbitals in the MPS in energy order, so that at any bond of the MPS, the left and right parts of the bath degrees of freedom always have touching energy spectra. This argument applies to an interacting model, too, because the bath is still noninteracting, and the Hubbard $U$ only reduces the chance for the impurity -- the only bridge via which the bath orbitals can indirectly hop to one another -- to be doubly occupied, thus reducing its bridging efficiency. The bath entanglement entropy of an interacting SIAM is therefore upper bounded by that of a noninteracting SIAM from this picture.

\subsection{Bath in chain geometry}

So far we have been working in the star geometry of the bath. Bath orbitals do not hop to each other directly. They only do so via the impurity. The diagonalization of bath orbitals in energy space leads to a logarithmic growth of entanglement entropy, according to the energy criterion in the previous section. In this section, we would like to emphasize again that the energy criterion is a sufficient but not necessary condition for the entropy to grow slowly. The example to give here is the evolution of the quenched SIAM in the chain geometry of the bath. The impurity is the head of the chain, which is directly connected to only one bath orbital, which in turn is connected to another bath orbital, and so on so forth. One can go from the star geometry to the chain geometry via Lanczos tridiagonalization starting from the impurity orbital, and from the chain back to the star by diagonalizing the bath. More details of the two geometries can be found in \cite{Wolf14}.

\begin{figure}
\includegraphics[width=0.95\columnwidth]{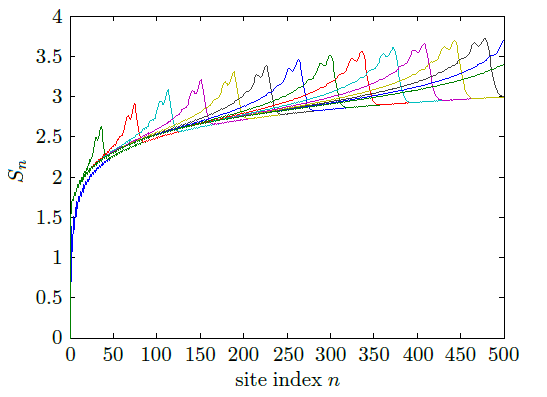}
\caption{(Color online) The entropy profiles at different times $Et=0$, $20$, $40$, $\ldots\,300$ in the chain geometry starting from $|0\rangle_d\otimes|\mathrm{FS}\rangle_\mathrm{bath}$ with $|\mathrm{FS}\rangle_\mathrm{bath}$ given in Fig.~\ref{fig:DOS}. Hubbard $U=0$ and impurity-bath coupling $V/E=0.25$. The number of bath orbitals $N=2000$.
\label{fig:entropy-profile}}
\end{figure}

\begin{figure}[b]
\includegraphics[width=0.95\columnwidth]{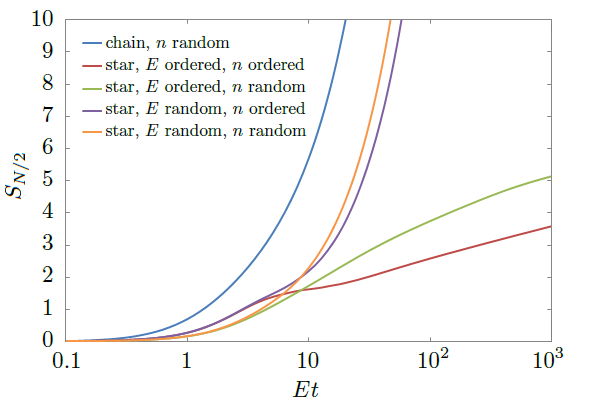}
\caption{(Color online) The growth of entropy of the noninteracting SIAM simulated in the chain and star geometries. The impurity-bath coupling $V/E=0.25$. The initial state is a product state with empty impurity and $0$-$1$ bath-orbital occupancies. ``$n$ ordered'' means the $n\leq N/2$ bath states are occupied and $n>N/2$ are empty. ``$E$ ordered'' in the star geometry means the bath-orbital energies are in ascending order of $n$. $S_{N/2}$ is the entanglement entropy between the $n\leq N/2$ bath orbitals and the rest of the system. The number of bath orbitals $N=2000$. The random results are averaged over 10 simulations.
\label{fig:entropy-geometries}}
\end{figure}

Starting from the initial state $|0\rangle_d\otimes|\mathrm{FS}\rangle_\mathrm{bath}$ with $|\mathrm{FS}\rangle_\mathrm{bath}$ being the same filled Fermi-sea state as in Fig.~\ref{fig:DOS} transformed to the chain geometry, the maximum entropy on the chain (the entanglement entropy between the left and right parts of the chain at the maximum entropy cut) is still found to grow logarithmically. Fig.~\ref{fig:entropy-profile} shows the result of a noninteracting calculation. The initial occupancies on the chain are spatially uniform. Every site has an occupancy of 0.5 per spin except the empty impurity. The entanglement entropy $S_n$ between sites $1,2,\ldots n$ and $n+1,\ldots N$ on the chain are then plotted in Fig.~\ref{fig:entropy-profile} as a function of $n$ at equal intervals of time. On top of the logarithmic background of $S_n$ of the equilibrium state $|\mathrm{FS}\rangle_\mathrm{bath}$, an entropy peak propagates like a soliton from the impurity down the chain at a speed $\sim E$. The maximum entanglement entropy (height of the peak) therefore increases with time logarithmically, even though there is no separation of energy spectrum on the chain, i.e., partition of the bath into different regions with different energies like in the star geometry.

Starting from an inverted half-filled Fermi-sea state with the highest energies initially occupied, the same logarithmic growth of entropy is seen due to particle-hole symmetry. But starting from a state with random initial occupancies in the star geometry, the entanglement profile in the chain geometry becomes prohibitively high ($\mathrm{max}(S_n)\propto N$) even at $t=0$. Also, a linear growth of entropy is seen starting from a product state in the chain geometry with $0$-$1$ random initial occupancies (see Fig.~\ref{fig:entropy-geometries}), while in the chain geometry, the entropy growth is still logarithmic starting from a random product state with $0$-$1$ occupancies. These results demonstrate that the logarithmic entropy growth in Fig.~\ref{fig:entropy-profile} is not guaranteed by the MPS basis of the Hamiltonian, but is due to the initial filled Fermi-sea state. For such a special initial state, the star geometry does not have a big advantage over the chain geometry, as they both give a logarithmic growth of maximum entanglement entropy. The benefit of the star geometry is its good behavior for more general initial states.

It is important to point out, as is shown in Fig.~\ref{fig:entropy-geometries}, that the star geometry alone does not guarantee a logarithmic entropy growth. The order of the bath orbitals in the MPS matters. The initial occupancies affect the transient growth of entropy, while the asymptotic entropy growth is determined by the ordering of the bath orbital energies.

\section{Double-impurity model \label{sec:double-impurity}}

\begin{figure}
\includegraphics[width=0.85\columnwidth]{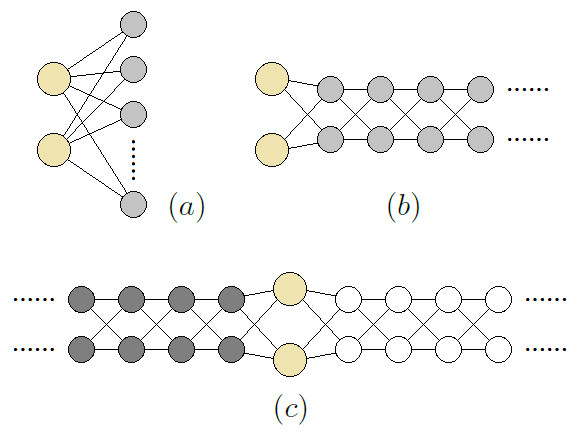}
\caption{(Color online) The general noninteracting double-impurity Anderson model in (a) the star geometry and (b) the chain geometry. Every orbital energy and every hopping line is an independent parameter. Panel (c) shows the double-impurity generalization of the two semi-infinite chain model in Fig.~\ref{fig:chain}.
\label{fig:DIAM}}
\end{figure}

In this section, we show that the logarithmic growth of entanglement entropy is not limited to the single-impurity Anderson model by doing a noninteracting simulation of a double-impurity Anderson model. The most general noninteracting double-impurity Anderson model can be pictorially represented in Fig.~\ref{fig:DIAM}. Fig.~\ref{fig:DIAM}a is in the basis in which the 2 impurity orbitals and all bath orbitals are diagonal, which is the double-impurity version of the star geometry. Fig.~\ref{fig:DIAM}b shows the double-impurity version of the chain geometry by Lanczos tridiagonalizing the star geometry in Fig.~\ref{fig:DIAM}a starting from the two impurities. One can also tridiagonalize the bath orbitals above and below the Fermi level separately (Fig.~\ref{fig:DIAM}c) to obtain the double-impurity generalization of the two semi-infinite chain model in Fig.~\ref{fig:chain}. Since the left and right semi-chains have touching energy spectra, the logarithmic growth of entropy is expected as a critical behavior between linear growth and saturation of entropy, as is discussed in Sec.~\ref{sec:results-entropy-analysis}.

\begin{figure}[b]
\includegraphics[width=0.9\columnwidth]{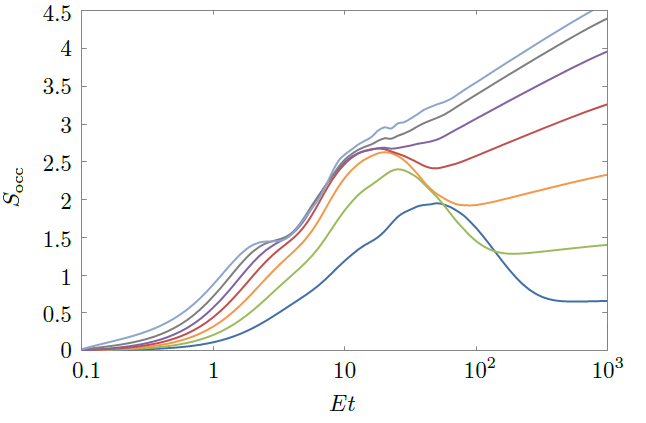}
\caption{(Color online) The logarithmic growth of entanglement entropy in a noninteracting double-impurity Anderson model. The bath DOS and filling are the same as Fig.~\ref{fig:DOS}. The two $d$ orbital energies $\epsilon_{d1,2}/E=\pm 0.2$ and $d_1-d_2$ hopping $V_{d1,d2}/E=0.15$. Both $d_1$ and $d_2$ are uniformly coupled to all bath orbitals with coupling amplitude $V/\sqrt{N}$ each. The coupling $V/E=0.1$, $0.15,\ldots,0.4$ from bottom to top. Number of bath orbitals $N=1000$.
\label{fig:entropy2}}
\end{figure}

Figure \ref{fig:entropy2} shows a sample result. We chose a half-filled bath with a semicircle DOS the same as Fig.~\ref{fig:DOS}, and put two $d$ orbitals at $\pm 0.2E$ ($E$ is the half band width) with $d-d$ hopping $0.15E$ to mimic typical crystal field splitting. The two $d$ orbitals are equally coupled to all bath orbitals. In the basis in which the two $d$ orbitals are diagonalized, their orbital energies are $\pm 0.25E$ and the original $d-d$ hopping makes the two $d$ orbitals now couple to the bath differently, which is more realistic. Then we plot the entanglement entropy $S_\mathrm{occ}$ between the initially occupied bath orbitals and the rest of the system.

The double-impurity model has a richer dynamics than SIAM.  Since both impurities are initially empty, the one below the Fermi level leaks a hole into the bath, leading to a short-term entropy peak. The steady-state growth of $S_\mathrm{occ}$ is still logarithmic, but the slope of $S_\mathrm{occ}$ v.s.~$\log t$ is not constant. This is because the two $d$ orbitals are not at the Fermi level (one is above and one is below). Their distances in energy to the Fermi level $|\epsilon_{d1,2}|$ relative to the impurity-bath coupling $V$ determine the slope, which approaches a maximum for the case of a $d$-orbital at the Fermi level ($\epsilon_d=0$) as $V$ gets large.

The logarithmic growth of entropy again shows that the quenched multi-impurity model is not exponentially hard in DMRG simulations, but is of only polynomial-time complexity. Whether the conclusion still holds for interacting models needs further investigation in DMRG, especially for those multi-impurity models with non-density-density (spin flipping and pair-hopping) terms, whose entanglement entropies need not be bounded by the corresponding noninteracting models.

\section{Conclusion \label{sec:conclusion}}

We have studied the growth of entanglement entropy in quenched Anderson impurity models. It is found that the growth of entropy is determined by the representation of the bath orbitals in the matrix product state (MPS). The Hubbard $U$ on the impurity orbital does not change the qualitative behavior of the steady-state growth of entanglement entropy of the bath MPS. The crucial feature controlling the growth of entanglement entropy is the overlap in energy of the density of states of the two parts of the maximum entanglement partition. In the star geometry of energy-ordered bath orbitals, the touching-spectra condition is satisfied at every bond, so the maximum bond dimension is power law in $t$. The power is upper bounded by the case of a half-filled $d$-orbital at the Fermi level and does not grow with the impurity-bath coupling, which allows a simulation of the long-time dynamics of the quenched impurity models in polynomial time. The conclusion is likely to generalize to multi-impurity models.

The growth of entanglement entropy of an interacting quantum system and the associated computational cost has been studied previously \cite{Prosen07,Pizorn14} in terms of the integrability of the quantum model. Our study looks at the problem from a different perspective. We focus on a special class of quantum models --- the impurity models --- and think of the growth of entanglement entropy among the bath orbitals. Because of the sparsity of interactions in the model, the entropy growth in the noninteracting bath is controlled by the energy partitioning of the bath and the localization of bath electrons to the energies they belong to. Since the new criterion of energy-partitioning the bath is not related in obvious ways to the integrability of the whole model (bath $\!+\!$ impurity), hopefully this new view of entropy growth of complexity can help us find new polynomial-time solvable models, parameter ranges, and/or special initial conditions that are not covered by the previously established integrability criterion.
\\

\noindent
\textbf{Acknowledgments:} We thank Dr.~Dante Kennes for helpful discussions. This research is supported by the
Department of Energy under grant DE-SC0012375.

\bibliography{SIAM1_refs}
\end{document}